\documentclass[a4paper]{jpconf}
\usepackage{graphicx}
\usepackage{array}

\begin{document}
\title{Measurement of neutrino interactions in gaseous argon with T2K}

\author{L Koch}
\address{for the T2K collaboration}
\address{RWTH Aachen University, 52056 Aachen, Germany}

\ead{koch@physik.rwth-aachen.de}

\begin{abstract}
The T2K near-detector, ND280, employs three large argon gas TPCs (Time
Projection Chambers) for particle tracking and identification. The gas
inside the TPCs can be used as an active target to study the neutrino
interactions in great detail. The low density of the gas leads to very
low track energy thresholds, allowing the reconstruction of very low
momentum tracks, e.g.  protons with kinetic energies down to
$\mathcal{O}(1\,\mathrm{MeV})$. Since different nuclear interaction
models vary considerably in their predictions of those low momentum track
multiplicities, this makes neutrino interactions on gases a powerful probe
to test those models.
The TPCs operate with an argon-based gas mixture ($95\%$ by volume) and have
been exposed to the T2K neutrino beam since the beginning of the
experiment in 2010. Due to the low total mass of the gas, neutrino argon
interactions happen only rarely, compared to the surrounding
scintillator-based detectors. We expect about $600$ such events in the
recorded data so far (about $200$ in the fiducial volume).
We are able to separate those events from the background and thus
demonstrate the viability of using gaseous argon as a target for a
neutrino beam. This enables us to do a cross-section measurement on
gaseous argon, the first measurement of this kind. All previous neutrino
cross-section measurements on argon were performed in liquid argon TPCs.
\end{abstract}

\section{T2K and the ND280 near detector}

The long-baseline neutrino experiment T2K uses the near detector ND280 -- 280 m
downstream from the accelerator's graphite target -- to measure the intensity
and composition of its neutrino beam at the source. It consists of
scintillation detectors, e.g. P0D, FGDs and ECALs, and three TPCs inside a
magnet yoke. The scintillators provide the target material for the neutrinos,
while the TPCs are instrumental in identifying the different particles
(see table~\ref{tab:tpc}) \cite{T2K}.

\begin{table}[b]
    \centering
    \caption[TPC technical data]{\label{tab:tpc}%
        TPC technical data
    }
    \vspace{1em}
    \begin{tabular}{m{0.4\textwidth}m{0.6\textwidth}}
        \hline
        Gas mixture (by volume) & Ar (95\%), CF (3\%), iC4H10(2\%) \\
        Gas density & $\sim1.74$ g/l (varies with T,p) \\
        Fiducial volume & $\sim 3 \times 1\,\mathrm{m}^3$ \\
        Expected number of neutrino\newline
            interactions (in FV) in current data\newline
            ($\sim 7\times10^{20}$ protons on target each)
        & $\sim600$ $(\sim200)$ in neutrino mode,\newline
            $\sim120$ $(\sim40)$ in anti-neutrino mode\\
        \hline
    \end{tabular}
\end{table}

\section{Nuclear interactions}

Nuclear effects, e.g. nucleon correlations and final state interactions (FSI),
are one of the dominant sources of systematic uncertainties for oscillation
analyses. They can produce low-momentum particles that are invisible to the
detectors and thus lead to a misreconstruction of the neutrinos' energy. Since
these low-momentum particles currently cannot be measured, the influence of the
nuclear effects has to be modeled and simulated.

\section{Neutrino gas interactions inside the TPCs}

The gas inside the TPCs can be used as an active target to study the neutrino
interactions in great detail. The low density of the gas leads to very low
track energy thresholds, allowing the reconstruction of very low momentum
tracks, e.g. protons with kinetic energies down to
$\mathcal{O}(1\,\mathrm{MeV})$. To achieve this, a new 3D reconstruction
algorithm was implemented.

\section{Generator models}

\begin{figure}
    \centering
    \includegraphics[width=0.9\textwidth]{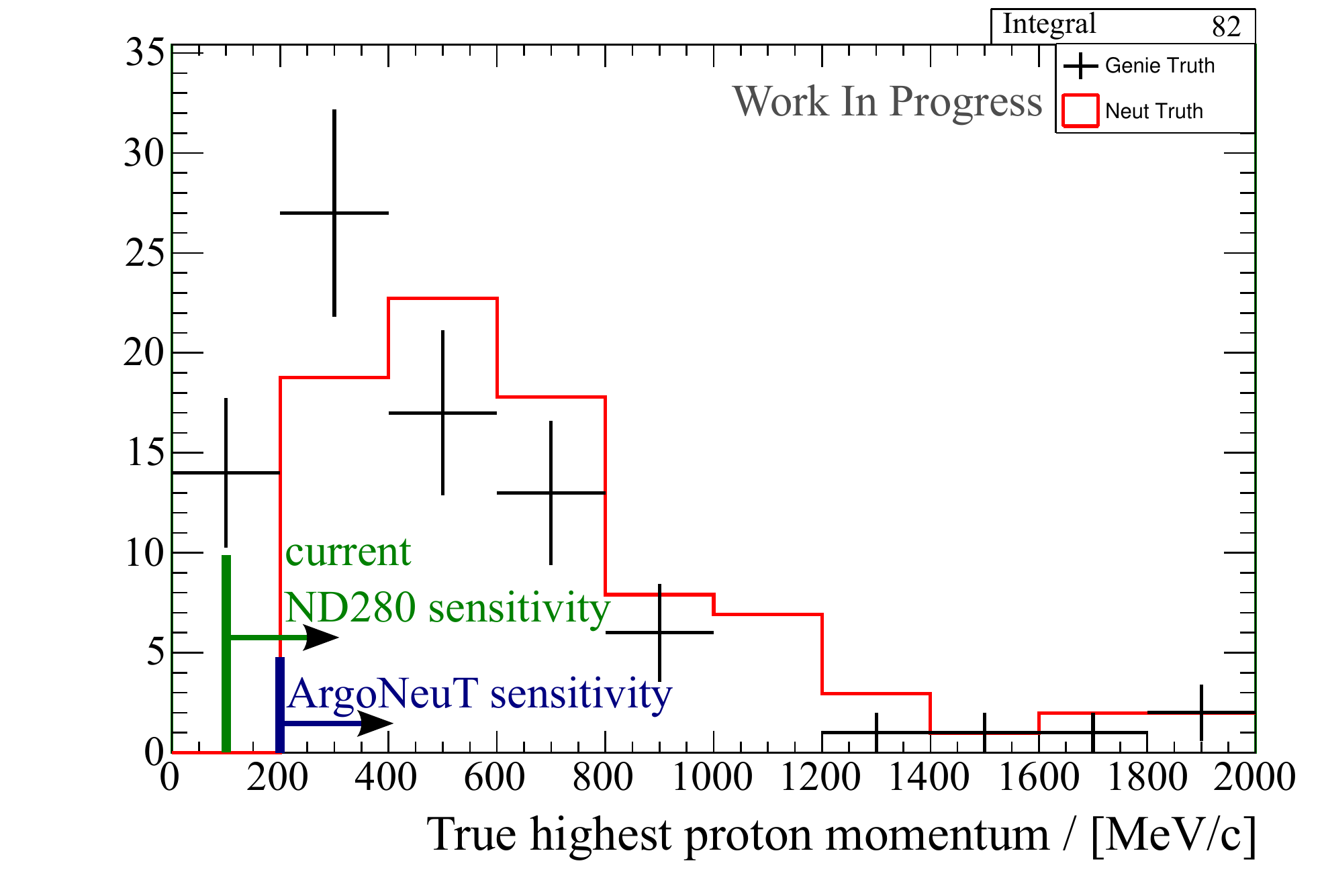}
    \vspace{-2em}
    \caption[Comparison of predicted proton moneta between event generators]{\label{fig:proton-mom}%
        Comparison of predicted proton momenta between event generators.
        The neutrino event generators Neut\cite{NEUT} and Genie\cite{GENIE} differ in their predictions of proton multiplicities,
        especially in the low-momentum region.
        The Neut data is area normalized to the Genie data.
        Previous cross-section measurements performed by the ArgoNeuT collaboration had a proton momentum threshold of 200\,MeV/c \cite{ARGONEUT}.
        The current threshold in the gaseous TPCs of ND280 lies at 100\,MeV/c and is expected to reduce with improvements of the reconstruction software.
    }
\end{figure}

Different nuclear interaction models have substantial differences in their
predictions of low momentum track kinematics (see figure~\ref{fig:proton-mom}).
This makes neutrino interactions on gases a powerful probe to test and constrain
them. ND280 is currently able to identify protons down to a momentum of
100\,MeV/c. This limit is expected to reduce to about 60\,MeV/c with improved
reconstruction and particle identification methods.

\section{Selection performance}

\begin{figure}
    \centering
    \includegraphics[width=0.9\textwidth]{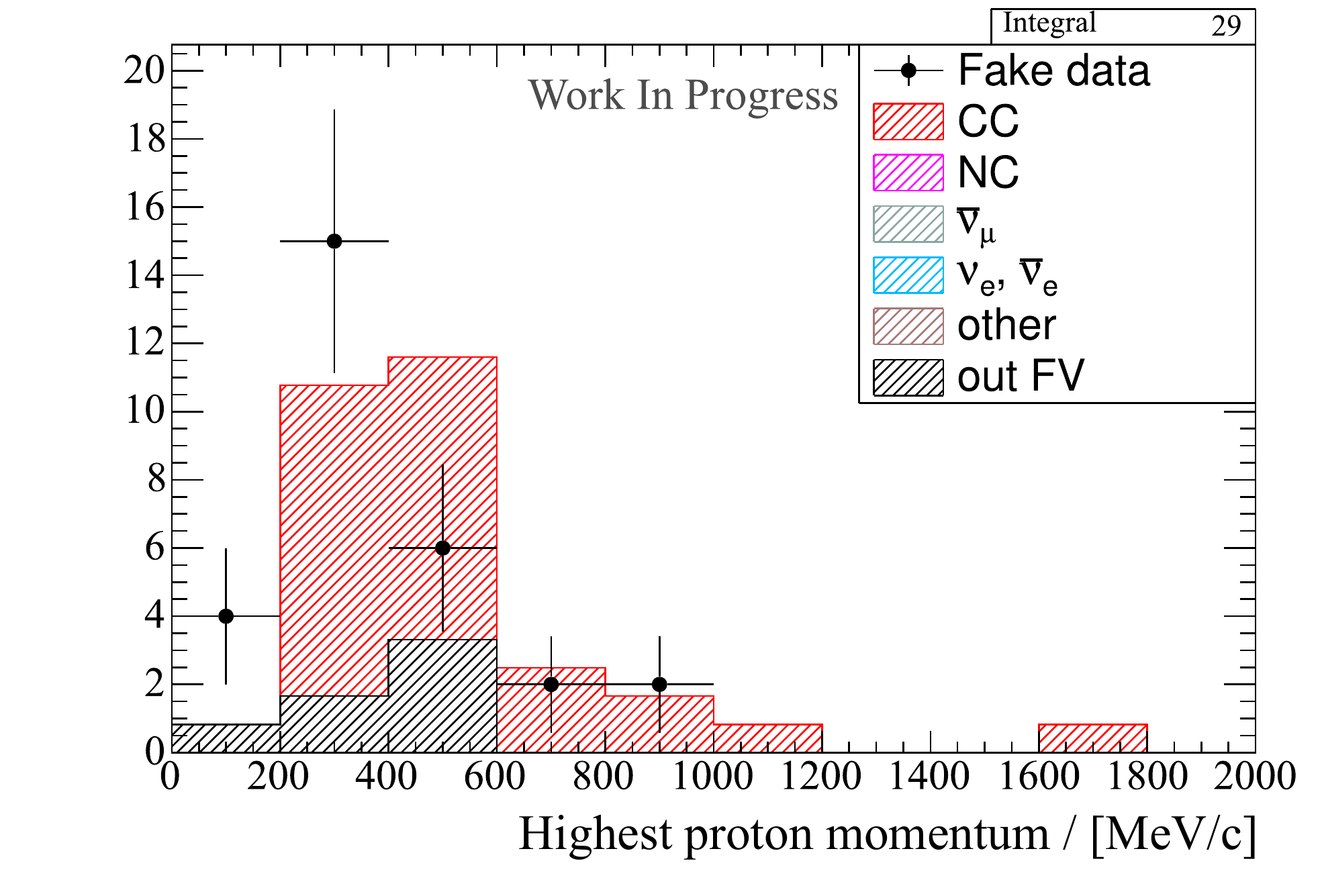}
    \vspace{-2em}
    \caption[Fake data distribution of proton momenta]{\label{fig:fake-data}%
        Fake data distribution of proton momenta.
        The fake data was generated with the event generator Genie.
        The Monte Carlo was generated with the event generator Neut.
        The error bars show statistical errors only.
        The Neut data is area normalized to the Genie data.
    }
\end{figure}

We perform a charged-current inclusive selection of neutrino interactions in the TPCs:
\[\nu_\mu + N \rightarrow \mu^- + X.\]
The achieved total purity is about 55\% and the efficiency about 45\%.
The numbers are identical for both the Neut and Genie generators,
within the statistic uncertainties of the Monte Carlo samples that have been looked at so far.
The selection looks for events with a muon starting in the TPC FV. The main
background consists of muons entering the TPCs from the outside and the
reconstruction misidentifying a vertex along its track.

A Monte Carlo study of the selection can be seen in figure~\ref{fig:fake-data}.
It compares the reconstruction and selection results between Genie~2.8.0 and Neut~5.3.2 Monte Carlo data.
The amount of data in each sample corresponds to $\sim 40\%$ of the amount of real data available for the final analysis.

\section{Conclusion}

The ND280 TPCs offer a unique opportunity to study nuclear effects in neutrino interactions with a very low detection threshold for the product particles.
We are able to separate the gas interaction events from the background and perform a charged-current inclusive selection.
This will enable us to test the kinematic predictions of different nuclear models.

\end{document}